\documentclass[pra,amsmath,amssymb,showpacs,superscriptaddress,twocolumn]{revtex4}

\usepackage{graphicx}
\usepackage{dcolumn}
\usepackage[hypertex]{hyperref}
\usepackage{bm,bbm}
\usepackage[usenames,dvipsnames]{color}
\usepackage{amsmath}
\usepackage{amsfonts}
\usepackage{amsthm}
\usepackage{amssymb}
\usepackage{mathrsfs}
\usepackage{subfigure}
\usepackage{ulem}

\newtheorem{theorem}{Theorem}

\newtheorem{corollary}{Corollary}

\begin{document}

\title{Quantum coherence in multipartite systems}

\author{Yao Yao}
\email{yaoyao@csrc.ac.cn}
\affiliation{Beijing Computational Science Research Center, Beijing, 100084, China}

\author{Xing Xiao}
\affiliation{Beijing Computational Science Research Center, Beijing, 100084, China}

\author{Li Ge}
\affiliation{Beijing Computational Science Research Center, Beijing, 100084, China}

\author{C. P. Sun}
\email{cpsun@csrc.ac.cn}
\affiliation{Beijing Computational Science Research Center, Beijing, 100084, China}
\affiliation{Synergetic Innovation Center of Quantum Information and Quantum Physics, University of Science and Technology of China,
Hefei, Anhui 230026, China}

\date{\today}

\begin{abstract}
Within the unified framework of exploiting the relative entropy as a distance measure of quantum correlations, we make explicit
the hierarchical structure of quantum coherence, quantum discord and quantum entanglement in multipartite systems. On this basis, we introduce
a new measure of quantum coherence, the \textit{basis-free} quantum coherence and prove that this quantity is exactly equivalent to
quantum discord. Furthermore, since the original relative entropy of coherence is a basis-dependent quantity, we investigate
the local and nonlocal unitary creation of quantum coherence, focusing on the two-qubit unitary gates. Intriguingly,
our results demonstrate that nonlocal unitary gates do not necessarily outperform the local unitary gates. Finally,
the additivity relationship of quantum coherence in tripartite systems is discussed in detail, where the strong subadditivity of
von Neumann entropy plays an essential role.
\end{abstract}


\maketitle
\section{INTRODUCTION}
In the context of quantum information theory, a distinct form of quantum resources corresponds to a specific
\textit{restriction} on the allowed quantum operations \cite{Coecke2014,Brandao2015}. Perhaps the best known
example along this line of thought is the quantum entanglement theory \cite{Vedral1997,Vedral1998,Plenio2007},
where the restricted set of operations is called local operations and classical communication (LOCC) \cite{Bennett1996a,Bennett1996b}.
To date, the theory of quantum entanglement has proven to be fruitful in various quantum information
tasks \cite{Horodecki2009} and directly inspired other resource theories of purity \cite{Horodecki2003}, the degree of superpositions \cite{Aberg2006},
thermodynamics \cite{Brandao2013,Gour2013}, quantum reference frames \cite{Gour2008,Marvian2013}, and the asymmetry of quantum states
\cite{Marvian2014}. The complete characterization of a particular resource theory mainly consists of three aspects: (i)
the unambiguous definition, (ii) the reasonable metrics, and (iii) the interconversions of quantum states under the
predetermined restrictions.

A recent successful application of quantum resource theory is the information-theoretic quantification of quantum coherence \cite{Baumgratz2014}.
Baumgratz \textit{et al.} proposed the basic notions of incoherent states, incoherent operations and a series of (axiomatic) necessary conditions any
measure of coherence should satisfy. Among all the potential metrics, the measures based on the $l_1$ norm and quantum relative entropy are highlighted.
This seminal work has triggered the community's interest in the definitions of other proper measures \cite{Girolami2014,Streltsov2015,Shao2015},
the freezing phenomenon \cite{Bromley2014}, the coherence transformations under incoherent operations \cite{Du2015},
and some further developments \cite{Xi2014,Pires2015,Bera2015,Singh2015,Yadin2015,Yuan2015}.

\begin{figure}[htbp]
\begin{center}
\includegraphics[width=0.40\textwidth ]{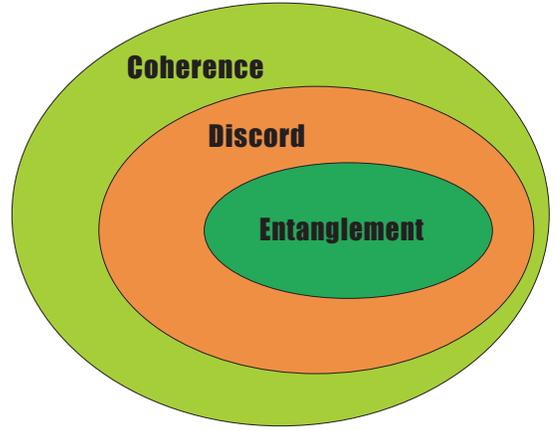}
\end{center}
\caption{(Color online) Venn diagram of different manifestations of quantum correlations present in composite quantum states.
}\label{fig1}
\end{figure}

However, it is worth noting that most of the related literatures have focused on a single qudit system and little attention has been paid to the
bipartite or multipartite systems \cite{Streltsov2015,Bromley2014}. In fact, the quantifications and classifications of quantum correlations
in multipartite systems are far from being settled up to now \cite{Horodecki2009,Modi2012}. In this work, we first establish the
hierarchical relationship of different manifestations of quantum correlations, on the basis of quantum relative entropy (see Fig. \ref{fig1}).
Furthermore, we pursue the answers to the following important issues:

$\bullet$ What is the exact relationship between quantum coherence with other measures of quantum correlations, such as quantum entanglement
or quantum discord ? Here we introduce the notion of \textit{basis-free} quantum coherence and prove this quantity is equivalent to quantum discord.
This correspondence relation opens up a new way to interpret the interconversions between different measures of quantum correlations.

$\bullet$ By definition, quantum entanglement $\mathcal{E}$ and discord $\mathcal{D}$ remain invariant under product (local) unitary transformations,
that is \cite{Horodecki2009,Modi2012}
\begin{align}
\mathcal{E}(\rho_{AB})&=\mathcal{E}(U_A\otimes U_B\rho_{AB}U_A^\dagger\otimes U_B^\dagger),\\
\mathcal{D}(\rho_{AB})&=\mathcal{D}(U_A\otimes U_B\rho_{AB}U_A^\dagger\otimes U_B^\dagger).
\end{align}
However, since quantum coherence is a \textit{basis-dependent} quantity, even local unitary transformations (let alone nonlocal operations) can
increase quantum coherence in bipartite systems. Therefore, it is worth investigating the local and nonlocal unitary creation of quantum coherence.

$\bullet$ In multipartite systems, a natural question arises how the correlations in the total system are distributed among the distinct subsystems.
For instance, we wonder whether the following relation holds for any tripartite state $\rho_{ABC}$
\begin{equation}
\mathcal{C}(\rho_{ABC})\geq\mathcal{C}(\rho_{AB})+\mathcal{C}(\rho_{AC}),
\end{equation}
where $\mathcal{C}(\rho)$ is a proper measure of quantum coherence.

\section{The resource theory of quantum coherence}\label{sec2}
To characterize quantum coherence as a physical resource, we first need to identify the definitions of incoherent states
and incoherent operations \cite{Baumgratz2014}. In an $N$-partite system, the incoherent states can be represented as \cite{Streltsov2015,Bromley2014}
\begin{equation}
\delta=\sum_{\vec{k}}\delta_{\vec{k}}|\vec{k}\rangle\langle\vec{k}|,
\end{equation}
where $|\vec{k}\rangle=|k_1\rangle\otimes|k_2\rangle\cdots\otimes|k_N\rangle$ and $|k_i\rangle$ is a \textit{pre-fixed} local basis of the $i$th subsystem.
According to the assumption on whether the measurement outcomes are recorded or not, the incoherent completely positive and trace preserving (ICPTP)
quantum operations are categorized into the following two classes \cite{Baumgratz2014}.

$\bullet$ \textit{The non-selective ICPTP maps:}
\begin{equation}
\Phi_{\text{ICPTP}}(\rho)=\sum_nK_n\rho K_n^\dagger,
\end{equation}
where the incoherent Kraus operators fulfill the constraints $\sum_n K_n^\dagger K_n=\mathbbm{1}$ and $K_{n}\mathcal{I}K_{n}^{\dagger}\subset\mathcal{I}$
for all $n$, where $\mathcal{I}$ denotes the whole set of incoherent states.

$\bullet$ \textit{The selective ICPTP maps:} these operations distinguish themselves from the above class by recording the measurement results, i.e.,
the post-measurement state corresponding to the outcome $n$ and its probability of occurrence are given by
\begin{equation}
\rho_n={K_n\rho K_n^\dagger}/{p_n},\quad p_n=\text{tr}[K_n\rho K_n^\dagger].
\end{equation}

Equipped with the above-mentioned theoretical definitions, Baumgratz \textit{et al.} presented a series of necessary conditions that any reasonable
measure of coherence should satisfy, in line with the resource theory of entanglement \cite{Vedral1997,Vedral1998}.

$\bullet$ (C1) \, $\mathcal{C}(\rho)=0$ iff $\rho\subset\mathcal{I}$;

$\bullet$ (C2a) \, Monotonicity under non-selective ICPTP maps, i.e.,
$\mathcal{C}(\rho)\geq\mathcal{C}(\Phi_{\text{ICPTP}}(\rho))$;

$\bullet$ (C2b) \, Monotonicity under selective ICPTP maps, i.e.,
$\mathcal{C}(\rho)\geq\sum_np_n\mathcal{C}(\rho_n)$;

$\bullet$ (C3) \, Convexity, i.e., $\sum_n p_n\mathcal{C}(\varrho_n)\geq\mathcal{C}(\sum_np_n\varrho_n)$
for any set of states $\{\varrho_n\}$ and any probability distribution $\{p_n\}$.

To satisfy the axiomatic conditions (C1), (C2b) and (C3), Baumgratz \textit{et al.} introduced the measures of coherence based on
$l_1$ norm and quantum relative entropy \cite{Baumgratz2014} while Girolami proposed another one by resort to the skew information \cite{Girolami2014}.
However, recently Du \textit{et al.} argued that the measure of coherence based on the skew information is probably
more applicable as a measure of asymmetry of quantum states \cite{Du2015a}. In this work, we mainly focus on the relative entropy of
coherence
\begin{align}
\mathcal{C}(\rho)=\min_{\delta\subset\mathcal{I}}S(\rho||\delta)
=S(\rho_{\mathcal{I}})-S(\rho),
\end{align}
where $\rho_{\mathcal{I}}$ is the diagonal version of $\rho$, which only retains the diagonal elements of $\rho$.

Before moving forward, it is interesting to take a closer look at the incoherent Kraus operators, which play an
essential role in the definition of incoherent operations. Indeed, the requirement $K\mathcal{I}K^{\dagger}\subset\mathcal{I}$
(here we omit the subscript $n$ for simplicity) is a rather strong constraint on the operator $K$. The following theorem tells us
that the structure or configuration of $K$ is highly restricted.

\begin{theorem}
There exists at most one nonzero entry in every column of the incoherent Kraus operator $K$.
\label{T1}
\end{theorem}

\textit{Proof.} The constraint $K\mathcal{I}K^{\dagger}\subset\mathcal{I}$ indicates that the incoherent Kraus operator
$K$ maps an \textit{arbitrary} incoherent state $\delta_a$ to an incoherent state $\delta_b$. Let us denote the elements of
the matrix $K$ as $[K]_{ij}=k_{ij}$. Similarly we can also represent the incoherent state $\delta_a$ as $[\delta_a]_{ij}=a_i\delta_{ij}$,
where $\{a_i\}$ are the diagonal entries of $\delta_a$ and $\delta_{ij}$ is the Kronecker delta. Therefore, adopting the Einstein's convention, we have
\begin{align}
[K]_{ij}[\delta_a]_{jl}[K^{\dagger}]_{lm}&=k_{ij}a_j\delta_{jl}k^\ast_{ml}
=a_jk_{ij}k^\ast_{mj}.
\end{align}
By use of $[\delta_b]_{ij}=b_i\delta_{ij}$, further we obtain
\begin{align}
\sum_ja_jk_{ij}k^\ast_{mj}=b_i\delta_{im}.
\label{summation}
\end{align}
Note that when $i\neq m$, the left hand side of Eq. (\ref{summation}) equals zero and the \textit{arbitrariness} of $\delta_a$ (thus $\{a_j\}$)
participates at this stage. If we choose the vector $\vec{a}=\{a_j\}=\{1,0,\cdots,0\}$, we have
\begin{align}
k_{i1}k^\ast_{m1}=0, \forall \, i\neq m,
\end{align}
which exactly implies that there exists at most one nonzero entry in the \textit{first} column of $K$.
The same reasoning can be easily generalized to other columns by a proper choice of $\{a_j\}$. \hfill $\blacksquare$

From Theorem \ref{T1}, we can directly obtain the following useful corollary.

\begin{corollary}
If the incoherent Kraus operator $K\subset \mathcal{M}_{s,t}$, where $\mathcal{M}_{s,t}$ denote the $s$ by $t$ matrices,
then the number of possible structure of $K$ is $s^t$. Here a legal structure stands for a possible arrangement of
nonzero entries in the matrix.
\end{corollary}

For example, as for $3\times2$ or $3\times3$ incoherent Kraus operators, the number of possible structure is $3^2=9$
and $3^3=27$, which easily recovers the result in Ref. \cite{Shao2015}.

\section{Hierarchies of multipartite quantum correlations}\label{sec3}
From geometric point of view, any distance measure between quantum states may serve as a candidate
for quantifying different forms of quantum correlations. A significant example is the usage of quantum relative entropy
in quantum information theory \cite{Vedral2002}. In particular, Vedral \textit{et al.} first proposed the
relative entropy of entanglement \cite{Vedral1997,Vedral1998} while the relative entropy of discord was first introduced by Modi \textit{et al.} \cite{Modi2010}.
Compared with the relative entropy of coherence, one can list the following definitions
\begin{align}
\mathcal{E}(\rho)&=\min_{\delta\subset\mathcal{S}}S(\rho||\delta),\\
\mathcal{D}(\rho)&=\min_{\delta\subset\mathcal{CC}}S(\rho||\delta),\\
\mathcal{C}(\rho)&=\min_{\delta\subset\mathcal{I}}S(\rho||\delta),
\end{align}
where $\mathcal{S}$ and $\mathcal{CC}$ stand for the sets of separable states and classically correlated states \cite{Modi2010}, respectively.
Since the incoherent states are diagonal states defined in a pre-determined orthogonal basis, the inclusion of sets clearly appears
\begin{align}
\mathcal{I}\subset\mathcal{CC}\subset\mathcal{S}
\end{align}
Therefore, we are led to the following hierarchical relations (see Fig. \ref{fig1})
\begin{align}
\mathcal{C}(\rho)\geq \mathcal{D}(\rho)\geq \mathcal{E}(\rho),
\end{align}
which signifies that quantum coherence is a far more ubiquitous manifestation of quantum correlations.

\begin{figure}[htbp]
\begin{center}
\includegraphics[width=0.40\textwidth ]{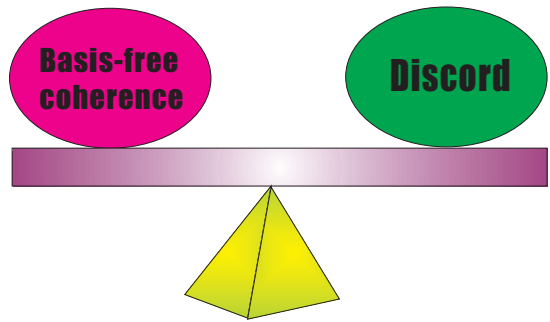}
\end{center}
\caption{(Color online) The equivalence between basis-independent quantum coherence and quantum discord.
}\label{fig2}
\end{figure}

On the other hand, it is worth emphasizing again that quantum coherence is a \textit{basis-dependent} quantity.
This pre-determined orthogonal basis $|\vec{k}\rangle=|k_1\rangle|k_2\rangle\cdots|k_N\rangle$
is a crucial premise when we refer to its computation or manipulation. However, under some circumstances
we are more inclined to deal with a \textit{basis-independent} quantity. Then a natural question arises whether
such a measure of quantum coherence can be defined. Here we first propose a basis-free measure of coherence
by the minimization over all local unitary transformations
\begin{align}
\mathcal{C}^{\text{free}}(\rho)=\min_{\vec{U}}\mathcal{C}(\vec{U}\rho\vec{U}^\dagger),
\end{align}
where $\vec{U}=U_1\otimes U_2 \cdots \otimes U_N$ possesses a local product structure.
The next theorem tells us that this quantity is \textit{exactly} equivalent to the relative entropy of discord.

\begin{theorem}
The basis-free quantum coherence $\mathcal{C}^{\text{free}}(\rho)$ is equal to $\mathcal{D}(\rho)$.
\end{theorem}

\textit{Proof.} With respect to a given basis $|\vec{k}\rangle$, the diagonal state $\rho_{\mathcal{I}}$ can be
represented as the completely decohered state of $\rho$
\begin{align}
\rho_{\mathcal{I}}=\sum_{\vec{k}}\langle\vec{k}|\rho|\vec{k}\rangle|\vec{k}\rangle\langle\vec{k}|.
\end{align}
Using this expression, we have
\begin{align}
\mathcal{C}^{\text{free}}(\rho)&=\min_{\vec{U}}\mathcal{C}(\vec{U}\rho\vec{U}^\dagger) \nonumber\\
&=\min_{\vec{U}}\left[S\left(\sum_{\vec{k}}\langle\vec{k}|\vec{U}\rho\vec{U}^\dagger|\vec{k}\rangle|\vec{k}\rangle\langle\vec{k}|\right)-S(\vec{U}\rho\vec{U}^\dagger)\right] \nonumber\\
&=\min_{\mathcal{B}(\vec{k})=\vec{U}^\dagger|\vec{k}\rangle}S\left(\sum_{\vec{k}}\langle\mathcal{B}(\vec{k})|\rho|\mathcal{B}(\vec{k})\rangle|\vec{k}\rangle\langle\vec{k}|\right)-S(\rho) \nonumber\\
&=\min_{\mathcal{B}(\vec{k})}H\left(\{|\mathcal{B}(\vec{k})\rangle\}\right)-S(\rho) \nonumber\\
&=\mathcal{D}(\rho),
\end{align}
where $\{|\mathcal{B}(\vec{k})\rangle=\vec{U}^\dagger|\vec{k}\rangle\}$ is a local orthogonal basis and $H(\{|\mathcal{B}(\vec{k})\rangle\})=-\sum_{\vec{k}}\langle\mathcal{B}(\vec{k})|\rho|\mathcal{B}(\vec{k})\rangle\log\langle\mathcal{B}(\vec{k})|\rho|\mathcal{B}(\vec{k})\rangle$.
In the derivation we have used the unitary invariance of von Neumann entropy and the results in Ref. \cite{Modi2010}. \hfill $\blacksquare$

This one-to-one correspondence builds a new bridge between quantum coherence and other forms of correlations and
opens up a new way to interpret the physical phenomena of quantum coherence (see Fig. \ref{fig2}). For example,

$\bullet$ It has been pointed out that nonclassical multipartite correlations (relative entropy of discord)
can be activated into distillable bipartite entanglement \cite{Piani2011,Streltsov2011}.
From the corresponding relationship between $\mathcal{C}^{\text{free}}(\rho)$ and $\mathcal{D}(\rho)$,
it is reasonable to conjecture that quantum coherence can also be considered as a resource for entanglement creation
and recently Streltsov \textit{et al.} have proved it is the case \cite{Streltsov2015}.

$\bullet$ For quantum discord, a freezing phenomenon occurs under certain initial conditions, especially when the underlying
system is subject to the environmental noise \cite{Modi2012}. From the equivalence relation between $\mathcal{C}^{\text{free}}(\rho)$ and $\mathcal{D}(\rho)$,
the same phenomenon may appear for quantum coherence \cite{Bromley2014}. In fact, Cianciaruso \textit{et al.} have
demonstrated that the freezing phenomenon of geometric quantum correlations is independent of the adopted distance measure
and thus universal \cite{Cianciaruso2014}.

\section{Local and nonlocal unitary creation of quantum coherence}\label{sec4}
From the definition of $\mathcal{C}^{\text{free}}(\rho)$ and its equivalence to the relative entropy of discord, we can easily
find that quantum coherence can be created by local (and nonlocal) unitary transformations. In this section, we concentrate on
the creation of quantum coherence in the context of two-qubit unitary gates. More precisely, we aim to evaluate the optimal
creation of coherence under specified types of unitary operators for a \textit{given} incoherent state, that is
\begin{align}
\mathcal{C}^{\text{opt}}=\max_{U_{AB}}\mathcal{C}(U_{AB}\delta_{\mathcal{I}}U_{AB}^\dagger),
\end{align}
where $\delta_{\mathcal{I}}=\mathbf{diag}\{\delta_1,\delta_2,\delta_3,\delta_4\}$ in the computational basis $\{|00\rangle,|01\rangle,|10\rangle,|11\rangle\}$
and $U_{AB}$ may be faced with some restrictions on its structure. Without loss of generality, we can arrange $\delta_i$ in ascending order
(that is, $0\leq\delta_1\leq\delta_2\leq\delta_3\leq\delta_4\leq1$). In the following, we mainly focus on three different types of two-qubit gates.

$\bullet$ \textit{One-side unitary operator $U_{AB}=U_A\otimes\mathbbm{1}_{B}$}. Using again the unitary invariance of
von Neumann entropy, $\mathcal{C}^{\text{opt}}$ can be rewritten as
\begin{align}
\mathcal{C}^{\text{opt}}=\max_{U_{AB}}S(\rho_{\mathcal{I}})-S(\delta_{\mathcal{I}}),
\label{copt}
\end{align}
where $\rho=U_{AB}\delta_{\mathcal{I}}U_{AB}^\dagger$. From Eq. (\ref{copt}), we only need to evaluate the four
diagonal entries of $\rho$. In the meantime, we can parametrize the general one-qubit unitary operator as
\begin{align}
U_A=e^{i\varphi}
\left(\begin{array}{cc}
a & b \\
-b^\ast & a^\ast
\end{array}\right)
\end{align}
where $|a|^2+|b|^2=1$. Note that in fact the overall phase $\varphi$ is irrelevant in our discussion, so we set $\varphi=0$.
For the sake of simplicity, we only present the four diagonal elements $\rho_{jj}$ of $\rho=U_A\otimes\mathbbm{1}_{B}\delta_{\mathcal{I}}U_A^\dagger\otimes\mathbbm{1}_{B}$
\begin{align}
\rho_{11}=|a|^2\delta_1+|b|^2\delta_3,\, \rho_{22}=|a|^2\delta_2+|b|^2\delta_4,\nonumber\\
\rho_{33}=|a|^2\delta_3+|b|^2\delta_1,\, \rho_{44}=|a|^2\delta_4+|b|^2\delta_2.
\end{align}

Remarkably, the effective role of $U_{AB}=U_A\otimes\mathbbm{1}_{B}$ is a \textit{mixture} of the diagonal elements of $\delta_{\mathcal{I}}$.
To find the optimal value $\mathcal{C}^{\text{opt}}$, we define the entropy function $F(|a|^2)=S(\rho_{\mathcal{I}})=\sum_j-\rho_{jj}\log_2\rho_{jj}$.
After simplification, the first derivative of $F(|a|^2)$ is
\begin{align}
\frac{\partial F(|a|^2)}{\partial(|a|^2)}&=(\delta_3-\delta_1)\log_2\frac{\delta_3+|a|^2(\delta_1-\delta_3)}{\delta_1+|a|^2(\delta_3-\delta_1)} \nonumber\\
&+(\delta_4-\delta_2)\log_2\frac{\delta_4+|a|^2(\delta_2-\delta_4)}{\delta_2+|a|^2(\delta_4-\delta_2)}.
\end{align}
With this expression and the ordering $\delta_1\leq\delta_2\leq\delta_3\leq\delta_4$, it is evident that the function $F(|a|^2)$ is monotonically increasing
for $|a|^2\in[0,1/2]$, while it is monotonically decreasing for $|a|^2\in[1/2,1]$. Therefore, when $|a|^2=1/2$ we arrive at the optimal value
\begin{align}
\mathcal{C}^{\text{opt}}_1=&-(\delta_1+\delta_3)\log_2(\delta_1+\delta_3)-(\delta_2+\delta_4)\log_2(\delta_2+\delta_4) \nonumber\\
&+1-\sum_{i=1}^4\delta_i\log_2\delta_i.
\end{align}
In order to distinguish it from the Hadamard gate $H$ \cite{Nielsen}, we denote the optimal one-qubit unitary operator as
(up to a global phase)
\begin{align}
U_A=\widetilde{H}_A=\frac{1}{\sqrt{2}}
\left(\begin{array}{cc}
1 & 1 \\
-1 & 1
\end{array}\right)
=|-\rangle\langle0|+|+\rangle\langle1|,
\end{align}
where $|\pm\rangle=(|0\rangle\pm|1\rangle)/\sqrt{2}$.

$\bullet$ \textit{Two-side unitary operator $U_{AB}=U_A\otimes U_{B}$}. Following a similar procedure, we first parametrize
the one-qubit unitary operators
\begin{align}
U_A=
\left(\begin{array}{cc}
a & b \\
-b^\ast & a^\ast
\end{array}\right),
U_B=
\left(\begin{array}{cc}
c & d \\
-d^\ast & c^\ast
\end{array}\right),
\end{align}
where $|a|^2+|b|^2=|c|^2+|d|^2=1$. Here we also provide the four diagonal entries of $\rho=U_A\otimes U_{B}\delta_{\mathcal{I}}U_A^\dagger\otimes U_{B}^\dagger$
\begin{align}
\rho_{11}&=|a|^2|c|^2\delta_1+|a|^2|d|^2\delta_2+|b|^2|c|^2\delta_3+|b|^2|d|^2\delta_4,\nonumber\\
\rho_{22}&=|a|^2|d|^2\delta_1+|a|^2|c|^2\delta_2+|b|^2|d|^2\delta_3+|b|^2|c|^2\delta_4,\nonumber\\
\rho_{33}&=|b|^2|c|^2\delta_1+|b|^2|d|^2\delta_2+|a|^2|c|^2\delta_3+|a|^2|d|^2\delta_4,\nonumber\\
\rho_{44}&=|b|^2|d|^2\delta_1+|b|^2|c|^2\delta_2+|a|^2|d|^2\delta_3+|a|^2|c|^2\delta_4.
\end{align}

Intriguingly, now the effective role of $U_{AB}=U_A\otimes U_{B}$ is a more thorough mixing of the diagonal elements of $\delta_{\mathcal{I}}$.
Instead of carrying out a similar analysis as in the first case, we can obtain the optimal value intuitively by noting that $S(\rho_{\mathcal{I}})\leq2$
for all $\rho$. Therefore, if $|a|^2=|b|^2=|c|^2=|d|^2=1/2$, we get the optimal value
\begin{align}
\mathcal{C}^{\text{opt}}_2=2-\sum_{i=1}^4\delta_i\log_2\delta_i,
\end{align}
where the constraint $\sum_i\delta_i=1$ is applied. By the concavity of function $-x\log_2x$, it is easy to verify that
$\mathcal{C}^{\text{opt}}_2\geq\mathcal{C}^{\text{opt}}_1\geq0$, which means the two-side local unitary operator performs better
in the creation of coherence. Moveover, the optimal unitary operator is $U_{AB}=\widetilde{H}_A\otimes\widetilde{H}_{B}$.

$\bullet$ \textit{The kernel of nonlocal unitary operator $U_d$}. In fact, any two-qubit unitary gate can be decomposed
in Cartan form \cite{Kraus2001,Khaneja2001,Zhang2003}
\begin{align}
U_{AB}=(X_A\otimes X_B)U_d(Y_A\otimes Y_B),
\end{align}
where $X_A$, $X_B$, $Y_A$ and $Y_B$ are single-qubit unitary operators and the bipartite nonlocal unitary \textit{kernel} $U_d$ has the form
\begin{align}
U_{d}(\vec{c})=\exp\left(-i\sum_{j=1,2,3}c_j\sigma_j\otimes\sigma_j\right).
\end{align}
Here $\sigma_j$ are standard Pauli operators and $\vec{c}=(c_1,c_2,c_3)$ is a real vector satisfying \cite{Kraus2001,Khaneja2001,Zhang2003}
\begin{align}
0\leq |c_3|\leq c_2\leq c_1\leq \pi/4.
\end{align}

Indeed, we should point out that coherence creation under arbitrary two-qubit gate can not be reduced to the problem where only $U_d$ is
taken into consideration, since we have already demonstrated that quantum coherence can be increased by local product unitary operators.
However, compared with the two previous cases, it is of significance to investigate the effect of the nonlocal kernel separately.
For clarity, we present the detailed discussion and some further expansion in the Appendix and the optimal $U_d$ is the kernel of the CNOT gate \cite{Zhang2003,Rezakhani2004}
\begin{align}
U_d(\pi/4,0,0)=\frac{1}{\sqrt{2}}\left(\mathbbm{1}-\sigma_1\otimes\sigma_1\right).
\end{align}
The corresponding optimal value is
\begin{align}
\mathcal{C}^{\text{opt}}_3=&-(\delta_1+\delta_4)\log_2(\delta_1+\delta_4)-(\delta_2+\delta_3)\log_2(\delta_2+\delta_3) \nonumber\\
&+1-\sum_{i=1}^4\delta_i\log_2\delta_i.
\end{align}
From the concavity of von Neumann entropy and the majorization theory \cite{Nielsen}, we have the ordering
\begin{align}
\mathcal{C}^{\text{opt}}_1\leq\mathcal{C}^{\text{opt}}_3\leq\mathcal{C}^{\text{opt}}_2,
\end{align}
which implies that the nonlocal kernel alone does not necessarily outperform the local product unitary operators concerning the creation of coherence.

To intuitively understand the physics behind these results, we notice the effect of the gate $\widetilde{H}$
\begin{align}
\widetilde{H}|0\rangle=|-\rangle,\,\widetilde{H}|1\rangle=|+\rangle,
\end{align}
that is, $\widetilde{H}$ transforms the computational basis states into the \textit{maximally coherent states} \cite{Baumgratz2014}.
More generally, if we apply $\widetilde{H}^{\otimes n}$ on $|1\rangle^{\otimes n}$, we have
\begin{align}
\widetilde{H}|1\rangle\otimes\widetilde{H}|1\rangle\otimes\cdots\otimes\widetilde{H}|1\rangle=\frac{1}{\sqrt{2^n}}\sum_{j=1}^{2^n-1}|j\rangle,
\end{align}
which is a $2^n$-dimensional maximally coherent states. In this particular sense, $\widetilde{H}$ (or the Hadamard gate $H$) can be regarded
as a \textit{maximally coherent operator}. In contrast, the CNOT gate is more inclined to create entanglement by noting that
\begin{align}
U_d(\pi/4,0,0)|00\rangle=\frac{|00\rangle-i|11\rangle}{\sqrt{2}},\nonumber\\
U_d(\pi/4,0,0)|01\rangle=\frac{|01\rangle-i|10\rangle}{\sqrt{2}},\nonumber\\
U_d(\pi/4,0,0)|10\rangle=\frac{|10\rangle-i|01\rangle}{\sqrt{2}},\nonumber\\
U_d(\pi/4,0,0)|11\rangle=\frac{|11\rangle-i|00\rangle}{\sqrt{2}},
\end{align}
which indicates that the nonlocal kernel of the CNOT gate transforms a fully separable basis into a maximally entangled basis \cite{Rezakhani2004}.
In fact, an arbitrary (two-qubit) incoherent states $\delta_{\mathcal{I}}$ can be converted to a \textit{Bell-diagonal-like} state by
the CNOT gate.

\section{Additivity relation of quantum coherence in tripartite systems}\label{sec5}
In this section, we discuss the additivity relation of quantum coherence in the tripartite scenario.
Here the \textit{additivity relation} describes how quantum coherence is distributed among the subsystems \cite{Yang2013}.
In particular, we wonder whether the tripartite coherence is equal to or greater than the sum of the
bipartite coherences, that is, whether the following inequality holds or not
\begin{equation}
\mathcal{C}(\rho_{ABC})\geq\mathcal{C}(\rho_{AB})+\mathcal{C}(\rho_{AC}),
\label{Add-relation}
\end{equation}
where $\rho_{AB}=\textrm{Tr}_C(\rho_{ABC})$ and $\rho_{AC}=\textrm{Tr}_B(\rho_{ABC})$. First, we
present two important class of states which are in favor of the inequality (\ref{Add-relation}).

$\bullet$ \textit{The generalized GHZ states $|\psi\rangle=\alpha|000\rangle+\beta|111\rangle$}.
In the computational basis, it is easy to verify that $\mathcal{C}(\rho_{ABC})=-|\alpha|^2\log_2|\alpha|^2-|\beta|^2\log_2|\beta|^2$
and $\mathcal{C}(\rho_{AB})=\mathcal{C}(\rho_{AC})=0$. Thus the inequality (\ref{Add-relation}) holds
in this case.

$\bullet$ \textit{The generalized W states $|\phi\rangle=\alpha|001\rangle+\beta|010\rangle+\gamma|100\rangle$}.
In the computational basis, we have
\begin{align}
\mathcal{C}(\rho_{AB})=&S(\rho_{\mathcal{I}}^{AB})-S(\rho_{AB})=S(\rho_{\mathcal{I}}^{AB})-S(\rho_{C})\nonumber\\
=&-|\beta|^2\log_2\frac{|\beta|^2}{|\beta|^2+|\gamma|^2}-|\gamma|^2\log_2\frac{|\gamma|^2}{|\beta|^2+|\gamma|^2},
\end{align}
where $\rho_{\mathcal{I}}^{X}$ is the diagonal version of $\rho_{X}$. Similarly, we also obtain
\begin{align}
\mathcal{C}(\rho_{AC})=&-|\alpha|^2\log_2\frac{|\alpha|^2}{|\alpha|^2+|\gamma|^2}-|\gamma|^2\log_2\frac{|\gamma|^2}{|\alpha|^2+|\gamma|^2},\nonumber\\
\mathcal{C}(\rho_{ABC})=&-|\alpha|^2\log_2|\alpha|^2-|\beta|^2\log_2|\beta|^2-|\gamma|^2\log_2|\gamma|^2.
\end{align}
Therefore, we have the inequality
\begin{align}
&\mathcal{C}(\rho_{AC})+\mathcal{C}(\rho_{AB})-\mathcal{C}(\rho_{ABC})\nonumber\\
=&(1-|\alpha|^2)\log_2(1-|\alpha|^2)+(1-|\beta|^2)\log_2(1-|\beta|^2)\nonumber\\
&-|\gamma|^2\log_2|\gamma|^2\leq0
\end{align}
To see this point, for a given $|\gamma|$, we define the function
\begin{align}
G(x)=x\log_2x+(a-x)\log_2(a-x),
\end{align}
where $x=1-|\alpha|^2\leq1$ and $a=1+|\gamma|^2\geq1$. It is easy to check that $G(x)$ is a \textit{convex}
function when $x\leq a$. Thus, the maximum value of $G(x)$ is reached at the boundary, that is, $\alpha=0$
or $\beta=0$.

The above evidences immediately tempt one to conjecture that the inequality (\ref{Add-relation}) holds for
any tripartite systems. Before attempting to construct or search a counterexample by
numerical simulation, the next theorem confirms that this conjecture is invalid by providing a
rather interesting class of states.

\begin{theorem}
There exists a class of states violating the additivity relation (\ref{Add-relation}), which satisfies strong subadditivity
of von Neumann entropy with equality.
\label{T3}
\end{theorem}

\textit{Proof.} For an arbitrary tripartite state $\rho_{ABC}$, we have
\begin{align}
&\mathcal{C}(\rho_{AC})+\mathcal{C}(\rho_{AB})-\mathcal{C}(\rho_{ABC})\nonumber\\
=&S(\rho_{\mathcal{I}}^{AB})-S(\rho_{AB})+S(\rho_{\mathcal{I}}^{AC})-S(\rho_{AC})\nonumber\\
&-S(\rho_{\mathcal{I}}^{ABC})+S(\rho_{ABC})\nonumber\\
=&[S(\rho_A)+S(\rho_{ABC})-S(\rho_{AB})-S(\rho_{AC})]\nonumber\\
&+[S(\rho_{\mathcal{I}}^{AB})+S(\rho_{\mathcal{I}}^{AC})-S(\rho_{\mathcal{I}}^{ABC})-S(\rho_{\mathcal{I}}^{A})]\nonumber\\
&+[S(\rho_{\mathcal{I}}^{A})-S(\rho_{A})]\nonumber\\
&=\Delta_1+\Delta_2+\Delta_3,
\end{align}
where $\Delta_1$, $\Delta_2$ and $\Delta_3$ represents the last three lines inside the square brackets respectively.
From the strong subadditivity of von Neumann entropy and the positivity of quantum coherence, we can determine
the sign of these three terms
\begin{align}
\Delta_1\leq0,\quad\Delta_2\geq0,\quad\Delta_3\geq0.
\end{align}
Therefore, when $\Delta_1=0$ we have the opposite inequality
\begin{equation}
\mathcal{C}(\rho_{AB})+\mathcal{C}(\rho_{AC})\geq\mathcal{C}(\rho_{ABC}).
\end{equation}
This completes the proof. \hfill $\blacksquare$

In fact, Hayden \textit{et al.} already presented an explicit characterization of the states which saturate the
strong subadditivity inequality for von Neumann entropy \cite{Hayden2004}. These states have the structure
\begin{align}
\rho_{ABC}=\bigoplus_jq_j\rho_{A^L_jB}\otimes\rho_{A^R_jC},
\end{align}
where $\{q_j\}$ is a  probability distribution and the Hilbert space of subsystem $A$ can be decomposed into a direct (orthogonal) sum of tensor products
\begin{align}
\mathcal{H}_A=\bigoplus_j\mathcal{H}_{A^L_j}\otimes\mathcal{H}_{A^R_j}.
\end{align}
In addition, we notice that the positivity of quantum discord was shown to be equivalent to the strong subadditivity of von Neumann
entropy \cite{Datta2010}. Theorem \ref{T3} tells us that the additivity relation in multipartite systems is also
closely related to the strong subadditivity of quantum entropy.

\section{CONCLUSIONS}\label{sec6}
In this work, we have systematically studied the quantum coherence in multipartite systems, employing
the quantum relative entropy as a distance measure. First, we characterize the structure of the incoherent
Kraus operators, which is a key ingredient in formulating the incoherent operations. Toward a unified view,
we present the hierarchical structure of quantum coherence, quantum discord and quantum entanglement
in multipartite systems. Remarkably, we propose the concept of basis-free quantum coherence and prove
that this quantity is exactly equivalent to the quantum discord. This one-to-one correspondence offers us
a new way to look at the interconversions between different types of quantum correlations. Moreover, we analytically
evaluate the optimal creations of quantum coherence for specific two-qubit unitary gates and the roles
of the Hadamard-like gate $\widetilde{H}$ and CNOT gate are highlighted. Finally, we explicitly figure out
the intrinsic connection between the additivity relation and the strong subadditivity of quantum entropy.

Within the framework of this work, there are several open questions to be addressed. (i) A detailed analysis
of the coherent power (capacity) of unitary operations is still missing (see the definition and discussion
in the Appendix). This aspect is of both theoretical and applied significance, since the creation and
maintenance of quantum coherence are a central problem in quantum communication and computation \cite{Nielsen};
(ii) Similar to the additivity relation discussed in this work, it is well know that
the monogamy or polygamy relations exist for quantum entanglement and discord \cite{Horodecki2009,Modi2012}.
For instance, we may check whether the following inequality holds for any tripartite states, in the spirit
of the seminal work by Coffman \textit{et al.} \cite{Coffman2000}
\begin{equation}
\mathcal{C}_{AB}+\mathcal{C}_{AC}\leq\mathcal{C}_{A(BC)}.
\end{equation}
Here the crux of this problem is how to appropriately define the quantum coherence for a bipartite partition.

\begin{acknowledgments}
This research is supported by the National Natural Science Foundation of China (Grant No. 11121403 and No. 11247006),
the National 973 program (Grants No. 2012CB922104 and No. 2014CB921403),
and the China Postdoctoral Science Foundation (Grant No. 2014M550598).
\end{acknowledgments}
%%%%%%%%%%%%%%%%%%%%%%%%%%%%%%%%%%%%%%%%%%%%%%%%%%%%%%%%%%%%%%%%%%%%%%%%%%%%%%%%%%%%%%%%%%%%%
\appendix
\section{Analysis of nonlocal unitary creation of quantum coherence}
In fact, the nonlocal kernel $U_d$ is diagonal in the magic basis \cite{Kraus2001}
\begin{align}
U_{d}=\sum_{k=1}^4e^{-i\lambda_k}|\Phi_k\rangle\langle\Phi_k|,
\end{align}
where the phases $\lambda_k$ are
\begin{align}
\lambda_1=c_1-c_2+c_3,\, \lambda_2=-c_1+c_2+c_3,\nonumber\\
\lambda_3=-c_1-c_2-c_3,\, \lambda_4=c_1+c_2-c_3.
\end{align}
Here the magic basis is
\begin{align}
|\Phi_1\rangle=|\Phi^+\rangle,\, |\Phi_2\rangle=-i|\Phi^-\rangle,\nonumber\\
|\Phi_3\rangle=|\Psi^+\rangle,\, |\Phi_4\rangle=-i|\Psi^+\rangle,
\end{align}
with $|\Phi^\pm\rangle=(|00\rangle+|11\rangle)/\sqrt{2}$ and $|\Psi^\pm\rangle=(|01\rangle+|10\rangle)/\sqrt{2}$.
Note that we always work in the standard computational basis, and then $U_d$ can be recast into the matrix form \cite{Rezakhani2004}
\begin{align}
U_d=
\left(\begin{array}{cccc}
e^{-ic_3}c^- & 0 & 0 & -ie^{-ic_3}s^- \\
0 & e^{ic_3}c^+ & -ie^{ic_3}s^+ & 0   \\
0 & -ie^{ic_3}s^+ & e^{ic_3}c^+ & 0   \\
-ie^{-ic_3}s^- & 0 & 0 & e^{-ic_3}c^-
\end{array}\right).
\end{align}
where $c^\pm=\cos(c_1\pm c_2)$ and $s^\pm=\sin(c_1\pm c_2)$.

Similar to the one-side case, we only need the four diagonal entries of $\rho=U_d\delta_{\mathcal{I}}U_d^\dagger$, that is
\begin{align}
\rho_{11}=(c^{-})^2\delta_1+(s^{-})^2\delta_4,\, \rho_{22}=(c^{+})^2\delta_2+(s^{+})^2\delta_3,\nonumber\\
\rho_{33}=(c^{+})^2\delta_3+(s^{+})^2\delta_2,\, \rho_{44}=(c^{-})^2\delta_4+(s^{-})^2\delta_1.
\end{align}
Since $(c^{\pm})^2+(s^{\pm})^2=1$, it is interesting to see that now the same reasoning in the one-side case can also
apply here. Therefore, the optimal condition is
\begin{align}
\cos^2(c_1\pm c_2)=\sin^2(c_1\pm c_2)=1/2,
\end{align}
which is equivalent to $c_1=\pi/4$ and $c_2=c_3=0$, under the constraint $0\leq |c_3|\leq c_2\leq c_1\leq \pi/4$.
The vector $\vec{c}=(\pi/4,0,0)$ exactly corresponds to the nonlocal kernel of the CNOT gate.

It is worth stressing that the definition of coherence creation here is not consistent with the so-called
entangling power (capacity) or discording power of a two-qubit unitary gate,
where the average or minimization is taken over the corresponding types of states \cite{Zanardi2000,Zanardi2001,Leifer2003,Linden2009,Galve2013}.
Along this line of thought, we can also define the \textit{coherent power (capacity)} of a gate $U_{AB}$ as
\begin{align}
\mathcal{CP}(U_{AB})=\max_{\delta\subset\mathcal{I}}\mathcal{C}(U_{AB}\delta U_{AB}^\dagger),
\end{align}
or more generally
\begin{align}
\mathcal{CP}(U_{AB})=\max_{\rho}[\mathcal{C}(U_{AB}\rho U_{AB}^\dagger)-\mathcal{C}(\rho)],
\end{align}
where $\rho$ may be restricted to a certain set. A systematic investigation of coherent power is underway.

%%%%%%%%%%%%%%%%%%%%%%%%%%%%%%%%%%%%%%%%%%%%%%%%%%%%%%%%%%%%%%%%%%%%%%%%%%%%%%%%%%%%%%%%%%%%%


\begin{thebibliography}{99}
\bibitem{Coecke2014} B. Coecke, T. Fritz, and R. W. Spekkens, arXiv:1409.5531.
\bibitem{Brandao2015} F. G. S. L. Brand\~{a}o and G. Gour, arXiv:1502.03149.
\bibitem{Vedral1997} V. Vedral, M. B. Plenio, M. A. Rippin, and P. L. Knight, Phys. Rev. Lett. \textbf{78}, 2275 (1997).
\bibitem{Vedral1998} V. Vedral and M. B. Plenio, Phys. Rev. A \textbf{57}, 1619 (1998).
\bibitem{Plenio2007} M. B. Plenio and S. Virmani, Quant. Inf. Comp. \textbf{7}, 1 (2007).
\bibitem{Bennett1996a} C. H. Bennett, H. J. Bernstein, S. Popescue and B. Schumacher, Phys. Rev. A \textbf{53}, 2046 (1996).
\bibitem{Bennett1996b} C. H. Bennett, D. P. DiVincenzo, J. A. Smolin and W. K. Wootters, Phys. Rev. A \textbf{54}, 3824 (1996).
\bibitem{Horodecki2009} R. Horodecki, P. Horodecki, M. Horodecki, and K. Horodecki, Rev. Mod. Phys. \textbf{81}, 865 (2009).
\bibitem{Horodecki2003} M. Horodecki, P. Horodecki, and J. Oppenheim, Phys. Rev. A \textbf{67}, 062104 (2003),
\bibitem{Aberg2006} J. {\AA}berg, arXiv:0612146.
\bibitem{Brandao2013} F. G. S. L. Brand\~{a}o, M. Horodecki, J. Oppenheim, J. M. Renes, and R. W. Spekkens, Phys. Rev. Lett. \textbf{111}, 250404 (2013).
\bibitem{Gour2013} G. Gour, M. P. M\"{u}ller, V. Narasimhachar, R. W. Spekkens, and N. Y. Halpern, arXiv:1309.6586.
\bibitem{Gour2008} G. Gour and R. W. Spekkens, New J. Phys. \textbf{10}, 033023 (2008).
\bibitem{Marvian2013} I. Marvian and R. W. Spekkens, New J. Phys. \textbf{15}, 033001 (2013).
\bibitem{Marvian2014} I. Marvian and R. W. Spekkens, Nat. Commun. \textbf{5}, 3821 (2014).

\bibitem{Baumgratz2014} T. Baumgratz, M. Cramer, and M. B. Plenio, Phys. Rev. Lett. \textbf{113}, 140401 (2014).
\bibitem{Girolami2014} D. Girolami, Phys. Rev. Lett. \textbf{113}, 170401 (2014).
\bibitem{Streltsov2015} A. Streltsov, U. Singh, H. S. Dhar, M. N. Bera, G. Adesso, arXiv:1502.05876.
\bibitem{Shao2015} L.-H. Shao, Z. Xi, H. Fan and Y. Li, Phys. Rev. A \textbf{91}, 042120 (2015).
\bibitem{Bromley2014} T. R. Bromley, M. Cianciaruso, and G. Adesso, Phys. Rev. Lett. \textbf{114}, 210401 (2015).
\bibitem{Du2015} S. Du, Z. Bai, and Y. Guo, Phys. Rev. A \textbf{91}, 052120 (2015) .
\bibitem{Xi2014} Z. Xi, Y. Li, and H. Fan, arXiv:1408.3194.
\bibitem{Pires2015} D. P. Pires, L. C. C\'{e}leri, and D. O. Soares-Pinto, Phys. Rev. A \textbf{91}, 042330 (2015).
\bibitem{Bera2015} M. N. Bera, T. Qureshi, M. A. Siddiqui, and A. K. Pati, arXiv:1503.02990.
\bibitem{Singh2015} U. Singh, M. N. Bera, H. S. Dhar, and A. K. Pati, Phys. Rev. A \textbf{91}, 052115 (2015).
\bibitem{Yadin2015} B. Yadin and V. Vedral, arXiv:1505.03792.
\bibitem{Yuan2015} X. Yuan, H. Zhou, Z. Cao, and X. Ma, arXiv:1505.04032.

\bibitem{Modi2012} K. Modi, A. Brodutch, H. Cable, T. Paterek, and V. Vedral, Rev. Mod. Phys. \textbf{84}, 1655 (2012).
\bibitem{Du2015a} S. Du and Z. Bai, Ann. Phys. \textbf{359}, 136 (2015).
\bibitem{Vedral2002} V. Vedral, Rev. Mod. Phys. \textbf{74}, 197 (2002).
\bibitem{Modi2010} K. Modi, T. Paterek, W. Son, V. Vedral, and M. Williamson, Phys. Rev. Lett. \textbf{104}, 080501 (2010).
\bibitem{Piani2011} M. Piani, S. Gharibian, G. Adesso, J. Calsamiglia, P. Horodecki, and A. Winter, Phys. Rev. Lett. \textbf{106}, 220403 (2011).
\bibitem{Streltsov2011} A. Streltsov, H. Kampermann, and D. Bru{\ss}, Phys. Rev. Lett. \textbf{106}, 160401 (2011).
\bibitem{Cianciaruso2014} M. Cianciaruso, T. R. Bromley, W. Roga, R. Lo Franco, and Gerardo Adesso, arXiv:1411.2978.
\bibitem{Nielsen} M. A. Nielsen and I. L. Chuang, \textit{Quantum Computation and Quantum Communication}
(Cambridge University Press, Cambridge, 2000).

\bibitem{Kraus2001} B. Kraus and J. Cirac, Phys. Rev. A \textbf{63}, 062309 (2001).
\bibitem{Khaneja2001} N. Khaneja, R. Brockett, and S. J. Glaser, Phys. Rev. A \textbf{63}, 032308 (2001).
\bibitem{Zhang2003} J. Zhang, J. Vala, S. Sastry, and K. B. Whaley, Phys. Rev. A \textbf{67}, 042313 (2003).
\bibitem{Rezakhani2004} A. T. Rezakhani, Phys. Rev. A \textbf{70}, 052313 (2004).
\bibitem{Yang2013} S. Yang, H. Jeong, and W. Son, Phys. Rev. A \textbf{87}, 052114 (2013).
\bibitem{Hayden2004} P. Hayden, R. Jozsa, D.Petz, and A. Winter, Commun. Math. Phys. \textbf{246}, 359 (2004).
\bibitem{Datta2010} A. Datta, arXiv:1003.5256.
\bibitem{Coffman2000} V. Coffman, J. Kundu, and W. K. Wootters, Phys. Rev. A \textbf{61}, 052306 (2000).

\bibitem{Zanardi2000} P. Zanardi, C. Zalka, and L. Faoro, Phys. Rev. A \textbf{62}, 030301(R)(2000).
\bibitem{Zanardi2001} P. Zanardi, Phys. Rev. A \textbf{63}, 040304(R) (2001).
\bibitem{Leifer2003} M. S. Leifer, L. Henderson, and N. Linden, Phys. Rev. A \textbf{67}, 012306 (2003).
\bibitem{Linden2009} N. Linden, J. A. Smolin, and A. Winter, Phys. Rev. Lett. \textbf{103}, 030501 (2009).
\bibitem{Galve2013} F. Galve, F. Plastina, M. G. A. Paris, and R. Zambrini, Phys. Rev. Lett. \textbf{110}, 010501 (2013).

\end{thebibliography}
\end{document}